\documentstyle{l-aa}            

\newcommand{\gsecdot}{^{\prime\prime}_{\raisebox{.6ex}{\hspace{.12em}.}}}
\newcommand{\gmindot}{^{\prime}_{\raisebox{.6ex}{\hspace{.12em}.}}}
\newcommand{\gfull}{^\circ}
\newcommand{\gfulldot}{^\circ_{\raisebox{.6ex}{\hspace{.05em}.}}}

\newcommand{\plm}{$\pm$}

\begin{document}
   \thesaurus{11         
              (05.01.1;  
               08.02.1;
               10.07.3)        }

   \title{ Astrometry of the globular cluster 47\,Tuc
        and possible optical identification of X-ray sources
                                     }
   \author  {M.~Geffert
           \inst{1}
      \and  M.~Auri\`ere
           \inst{2}
      \and  L. Koch-Miramond
           \inst{3}
             }

   \offprints{geffert@astro.uni-bonn.de}

   \institute{Sternwarte der Universit\"at Bonn, Auf dem H\"ugel 71, 
              D-53121 Bonn, Germany
              \and     Observatoire du Pic du Midi, UMR 5572, CNRS, 
                       F-65200 Bagn\`eres de Bigorre, France 
              \and     Service d'Astrophysique
                       Centre d'Etudes de Saclay,
                       F-91191 Gif sur Yvette Cedex, France}
   \date{Received date; accepted date}

   \maketitle
   \markboth {Geffert et al.: Astrometry of 47\,Tuc} {}

   \begin{abstract}
Positions of stars of  47\,Tuc 
have been derived by the use of photographic plates and  CCD frames 
(ESO La Silla)
and data from the Hubble Space Telescope (HST). 
The positions have been determined with respect
to the PPM, which is based on the FK5 system.
We have compared the positions of variable and blue stars
in the core of 47\,Tuc with those of the X-ray sources 
found by Hasinger et al.\, (1994). Taking into account
a possible constant shift of the X-ray positions of
up to 10 $\arcsec$, there are three different solutions
which would give identifications of four of the
central X-ray sources with blue stragglers and
variable stars. We discuss the nature of the X-ray 
sources and the different possible identifications. 
Since it is not possible to give an unique 
identification based on pure astrometric arguments, our
positions may be taken for future identifications 
based on additional astrophysical arguments or
coincident observations in different wavelengths.
An identification of the X-ray source No. 3 of
Hasinger et al.\, (1994) in the outer field of
47\,Tuc with a galaxy was found.

   \end{abstract}

\section{Introduction}

The dynamical evolution of dense stellar systems
like the cores of globular clusters may be significantly
influenced by binaries (e.g. Hut 1996 and references
therein).
In  a confrontation of theories with observations the
detection of possible tracers of binaries like
millisecond pulsars, X-ray binaries, cataclysmic
variables, or blue stragglers plays an important role.
One of the most promising objects for such a 
confrontation is the globular cluster
47\,Tuc, where more than 40 blue stragglers (Paresce et al.\, 1991,
Guhathakurta et al.\, 1992, De Marchi et al.\, 1993, Lauzeral 1993, 
Auri\`ere et al.\,1994),
11 millisecond pulsars (Manchester et al.\, 1991, Robinson et al.\, 1995),
2 cataclysmic variables (Paresce et al.\, 1992, Paresce \& De Marchi, 1994)
and 5 X-ray sources in the core (Hasinger et al.\, 1994),
have been detected. In order to investigate the physical
nature of these objects it is important to compare their stellar fluxes
at different wavelength domains, which requires an 
optical identification of the objects detected in other
wavelengths.  
Hasinger et al.\, (1994) have detected 15 X-ray sources in the
field of 47\,Tuc and discussed possible
optical  identifications of some of
them. Of peculiar interest are the core sources which are 
expected to really belong to the cluster, whereas some
of the external ones should belong to the field
(Johnston et al. 1996). 

X-ray globular cluster sources
are classically divided in two classes (Hertz and 
Grindlay, 1983). 

High luminosity ones are low-mass
X-ray binaries (LMXB). Two of them were optically identified
from the ground, in M15 (Auri\`ere et al.\, 1984; Ilovaisky
et al.\, 1993 and references herein) and in NGC 6712 
(Nieto et al.\, 1990, Auri\`ere \& Koch-Miramond 1992)
and confirmed by HST observations (Downes et al.\, 1996). A third 
one was resolved with the HST in NGC 6624 (King et al.\, 1993).
Recently, Deutsch et al.\, (1996) presented an UV
excess optical candidate for the luminous globular cluster 
X-ray source in NGC 1851.

Faint X-ray sources in globular clusters (so-called dim sources)
are expected to be cataclysmic binaries. In a recent review 
Johnston et al.\, (1996) pointed out the various spectral properties 
of these objects. They concluded that this group of X-ray sources 
consists of different types of sources.
At least in the globular cluster NGC 6397 Grindlay et al. (1995) 
conclusively identified three suspected optical counterparts (De Marchi \& 
Paresce 1994, Cool et al.\, 1995) as magnetic cataclysmic variables.
The case of the core of 47\,Tuc is 
puzzling since the Einstein X-ray source was found to 
be too faint to be a LMXB and too bright to be a 
cataclysmic variable (Auri\`ere et al.\, 1989). The
discovery of several sources in the 47\,Tuc core by
ROSAT has not solved the problem since the sources 
are in the same intensity range (Hasinger et al.\, 1994).
One of the best considered hypotheses is that these 
sources are transient LMXB in quiescence (Bailyn, 1995),
as first advocated by Verbunt et al.\, (1984).
The basic problem in the identification 
process is to determine the link between the optical  
and the X-ray  positions. Due to
the crowding in the centre of 47\,Tuc and due to
a possible shift of the X-ray positions (Hasinger et
al. 1994) the number of possible optical counterparts 
of the X-ray sources is high.
In this paper we perform new astrometric measurements of the stars 
in 47\,Tuc using groundbased (ESO La Silla) observations 
as well as HST data in order to look for
the identifications of the X-ray sources in 47\,Tuc.

\begin{table*}
\caption[]{The observations and their specifications}
\begin{flushleft}
\begin{tabular}{lrrrrrr}
\hline\noalign{\smallskip}
Date & Exposure & NE  &    Telescope & Detector & Passband & Scale \\
     & time    &     &      &          &    &       \\
\noalign{\smallskip}
\hline\noalign{\smallskip}
Plates: & & & & & \\
1988-08-08   & 3,5 min & 2  &  ESO/GPO  &  IIa-O  &    B  &  1mm=50$\arcsec$ \\
             &         &    &      &         &       &                  \\
CCD frames:  &         &           &         &       &                  \\
 1986-07-26  & 1,2 min & 4 &  ESO/2.2m &  GEC    & 
   U  &  1px=0$\gsecdot$26  \\
 1990-07-28  & 4,1,0.1 min & 3 &  ESO/NTT  &  TEK    & 
U,B,V  &  1px=0$\gsecdot$15  \\
 1990-07-29  & 5,1,0.2 min & 3 & ESO/NTT &  TEK    & 
  U,B,V  &  1px=0$\gsecdot$15  \\
\noalign{\smallskip}
\hline
\end{tabular}
\end{flushleft}
NE = Number of plates/exposures
\end{table*}

\section {Observations and reductions}

The complete observational material used in this work is given in 
Table 1.
Three steps were needed to link the rectangular HST 
positions to the PPM system (see Table 2).
The first step was a determination of positions
of 30 secondary reference stars around 47\,Tuc by the use
of photographic plates and 44 PPM stars.
In a second step, positions of 2683 stars 
in a  central field of 4\arcmin $\times$ 4\arcmin 
 ~of 47\,Tuc
were determined from CCD frames.
In a third step we have transformed the rectangular coordinates 
of the stars from the HST observations from
Guhathakurta et al.\, (1992) and De~Marchi et al.\, (1993) 
into spherical ones.
Standard procedures as described
e.g. in Geffert et al.\, (1994) were used to measure and reduce
the photographic plates and CCD frames.
We obtained the root mean square (rms) of the deviations of our 
measurements from
the PPM catalogue of 200 milliarcseconds (mas) in each coordinate. 
An intercomparison
of the measurements of the two GPO plates yielded the rms of 150 mas.

DAOPHOT software (Stetson 1987) was used to determine the
rectangular positions of the stars from the CCD frames. 
The rms of the deviations between the catalogue of the 
secondary stars and the rectangular positions on the CCD 
frames was 80 mas. These differences were mainly caused by
the uncertainties of the secondary reference stars, since the deviations 
of x and y coordinates between the different CCD frames gave rms of 10 mas.
For the reduction of the CCD-frames also third order polynomials
of the rectangular coordinates x and y had to be taken into account.
As seen in Table 2 the main uncertainty of our final positions is due
to the determination of the positions with respect to the PPM
system which amounts to 100 mas.

\begin{table*}
\caption[]{The successive steps in our reduction}
\begin{flushleft}
\begin{tabular}{lrrrrrr}
\hline\noalign{\smallskip}
 &  & Reference & Reference & Target & Target & Transform.  \\
   &  &  Stars        &  Field       &    Stars  & Field & Uncertainty   \\
\noalign{\smallskip}
\hline\noalign{\smallskip}
STEP I & Phot. Plates &  44 & 2 $\gfull$ $\times$ 2 $\gfull$ & 30 & 
    2$\gmindot$6 $\times$ 2$\gmindot$6 & 0$\gsecdot$08 \\
STEP IIa & CCD frames & 30 & 2$\gmindot$6 $\times$ 2$\gmindot$6 & 2683 & 
          2$\gmindot$6 $\times$ 2$\gmindot$6 &  0$\gsecdot$01 \\
STEP IIIa & HST observations I & 33 & 2$\gmindot$6 $\times$ 2$\gmindot$6 & 
          57 & 0$\gmindot$7 $\times$ 0$\gmindot$7 & 0$\gsecdot10$ \\
STEP IIIb & HST observations II & 120 & 2$\gmindot$6 $\times$ 2$\gmindot6$ & 
          3567 & 0$\gmindot$3 $\times$ 0$\gmindot$3 & 0$\gsecdot01$ \\
\noalign{\smallskip}
\hline
\end{tabular}
\end{flushleft}
HST observations I are from Guhathakurta et al. (1992) \\
HST observations II are from De~Marchi et al. (1993) \\
\end{table*}

\begin{table*}
\caption[]{ Positions of variable and blue stars in the core of 
            47\,Tuc. The epoch of the positions is 1988. }
\begin{flushleft}
\begin{tabular}{rrrrrrrrr}
\hline\noalign{\smallskip}
AKO & P & GYSB & DPF & L & E & V  & $\alpha_{2000}$ & $\delta_{2000}$   \\
    &   &   &     &   &     & &  [hhmmss.sss]   &  [ddmmss.ss]     \\
\noalign{\smallskip}
\hline\noalign{\smallskip}
 &   & & & &  13     &     &  002401.376 & $-$720447.68 \\        
  &  & 32 & & 942 & &      &   002402.301 & $-$720456.11 \\
  &  & 84 & & 1028 & &     &   002402.996 & $-$720519.61 \\
 &   & & & &   9     &     &  002403.262 & $-$720426.69 \\
  &   & & & &   8    &     &  002403.448 & $-$720504.61 \\
 & 1 & 172 & 1596 & & &    &  002404.034 & $-$720450.91 \\
 &  & & 2213 & & &     1    &  002404.286 & $-$720457.68 \\
 &  & 206 & & 1200 & &     &  002404.484 & $-$720454.63 \\
6 & 2  & & 1984  & & &     &  002404.489 & $-$720454.72 \\
 & 3  & 253 & 2646 & & &   &  002404.886 & $-$720501.67 \\
9 &  & & 2059 & & 11 &     &  002404.946 & $-$720455.13 \\
 &  & & & 1273 & &         &  002405.046 & $-$720537.51 \\
 &  & 241 & & 1273 & &     &  002405.086 & $-$720537.49 \\
 &   & & & &   7     &     &  002405.244 & $-$720451.11 \\
 &  & 283 & & 1328 & &     &  002405.478 & $-$720515.52 \\
 & 4  & 302 & 1166 & 1334 & & & 002405.574 & $-$720444.93 \\
 &  5 & 299 & 1997 & & 6 &  &    002405.578 & $-$720453.67 \\
 & 6  & & 2772 & & &        &    002405.642 & $-$720502.57 \\
 & 7  & 312 & 2850 & & &    &    002405.700 & $-$720503.76 \\
 & 8  & & 1953 & & &        &    002405.759 & $-$720453.05 \\
 & 9  & & 2588 & & &        &    002405.946 & $-$720459.94 \\
 & 10  & & 1582 & 1418 & &  &     002405.990 & $-$720448.67 \\
 &    & &      & & &    2   &     002406.017 & $-$720455.85 \\
 & 11  & & 1581 & 1418 & &  &     002406.052 & $-$720448.60 \\
 & 12  & & 1757 & & &       &     002406.094 & $-$720450.51 \\
 &  & 381 & & 1430 & 1 &    &    002406.372 & $-$720437.05 \\
 &  & 382 & & 1447 & &      &  002406.494 & $-$720526.69 \\
 & 13  & 386 & 2064 & & &   &     002406.414 & $-$720453.68 \\
 & 15 & 389 & 1030 & & &    &    002406.415 & $-$720442.55 \\
 & 14  & & 1951 & & &       &     002406.446 & $-$720452.30 \\
 & 16  & & 1286 & & 12 &    &     002406.453 & $-$720445.16 \\
 & 17  & & 1929 & & &       &     002406.507 & $-$720452.01 \\
 & 18  & 398 & 1758 & 1450 & 3 & & 002406.532 & $-$720450.04 \\
 &  & 399 & & 1456 & &      &   002406.569 & $-$720502.90 \\
 & 21  & & 1206 & & &       &     002407.710 & $-$720442.98 \\
 &   & & & & 4       &     &  002407.822 & $-$720458.07 \\
 &  & 523 & & 1628 & &     &   002407.885 & $-$720514.00 \\
 & 19  & 534 & 2128 & & &   &     002407.925 & $-$720452.73 \\
 &   & & & &  10     &   &    002407.942 & $-$720501.64 \\
 & 20  & & 2732 & & &     &       002408.171 & $-$720459.31 \\
 &  & 572 & & 5280 & &     &   002408.573 & $-$720430.76 \\ 
 &  & 577 & & & &          &   002408.653 & $-$720510.28 \\
 &   & & & &   2     &     &  002409.070 & $-$720504.87 \\
 &  & 658 & & 1826 & &     &   002409.644 & $-$720444.92 \\
 &   & & & &   5     &     &  002410.028 & $-$720451.64 \\
 &  & & & 2089 & &         &   002412.135 & $-$720445.68 \\
 &  & 801 & &      & &     &   002412.253 & $-$720446.67 \\
 &  & 866 & &      & &     &   002413.587 & $-$720445.66 \\
 &  & 938 & &      & &     &   002415.794 & $-$720441.24 \\
\noalign{\smallskip}
\hline
\end{tabular}
\end{flushleft}
AKO = Number in Auri\`ere et al.\, (1989),
P = Number im Paresce et al.\, (1991) \\
GYSB = Number in Guhathakurta et al.\, (1992),
DPF = Number in De~Marchi et al.\, (1993) \\
L = Number in Lauzeral (1993),
E = Number in Edmonds et al.\, (1996) \\
V = Variable stars from Paresce et al.\, (1992) 
    and Paresce \& De~Marchi (1994) \\
\end{table*}

The transformation of the rectangular coordinates of
De~Marchi et al.\, (1993) and Guhathakurta et al.\, (1992) into
spherical ones suffers from the problem that, due to the
higher resolution, the
HST-observations show several stars at the place of one
star on the CCD frames.  Because of these 
difficulties we have
chosen a narrow search radius of 300 mas for the identification
of the HST stars corresponding to our CCD based catalogue. 
Still we  had the situation that
in some cases more than one star from the HST 
observations corresponds to one of our
catalogue stars. However, since we also excluded in the reduction
stars having a positional deviation of more than 3$\sigma$, the
accuracy of the transformation of the HST stars to our catalogue
- as seen below - was sufficient.
120 stars remained in the reduction for the transformation
from the rectangular HST coordinates of De Marchi et al.\, (1993)
into spherical ones in our catalogue system. The rms of the 
deviations between the HST coordinates and our catalogue
were 50 mas. The uncertainty of the transformation at the place of
one star was better than 10 mas. 

The situation is much worse with the positions of Guhathakurta
et al.\, (1992). We derived a rms of 280 mas in each coordinate
and an uncertainty of 100 mas at the place of one star. 
For stars in common with the catalogue of De~Marchi et al.\, (1993)
we took therefore preferentially the positions from their catalogue.

We made an astrometric catalogue of all candidates
for beeing the optical counterparts of the X-ray sources
in the core of 47\,Tuc. Our catalogue contains positions of
all known variable and blue stars (including the blue
stragglers).
Table 3 gives the cross-identifications and positions of these
stars. We have taken mainly  those positions which were
determined from the measurements of De~Marchi et al.\, (1993).
For those stars which were located outside the field of 
De~Marchi et al.\, (1993), we have taken preferentially the positions
of Guhathakurta et al.\, (1992), then those of Lauzeral (1993) and 
Edmonds et al.\, (1996).

\section{The X-ray positions in the field of 47\,Tuc}

Hasinger et al.\, (1994) list ROSAT X-ray positions from two 
observations in
1992 and 1993. While 15 sources were detected in the whole campaign, 
only 5 sources were visible in both observations.
 The small difference between the positions points
to a constant offset of the ROSAT positions, as was the case
in other observations (see remark in Hasinger et al.\, 1994 in
chapter 3). Since the internal errors of the second observation are 
smaller than the errors from the first one, we have transformed all 
positions to the system of the second observation by using
the five sources in common. The mean weighted deviations of the 
positions are (observation2$-$observation1) $-$0$\fs$22 \plm
0$\fs$15 seconds in right ascension and  $-$1$\gsecdot$2  \plm 
0$\gsecdot$8 in
declination. 

Since a constant offset of up to 5$\arcsec$ for the ROSAT 
positions is possible (Hasinger et al.\, 1994), the number 
of possible candidates for an optical counterpart 
in a central part of a globular cluster is quite large.
The optical identification of the X-ray sources in the central 
part is therefore not possible by a direct 
position to position comparison. In general the X-ray 
sources No. 7 and 9 are, with their small internal errors 
of the X-ray positions, the most promising candidates for
a determination of the  constant offset of the ROSAT positions.

\begin{table}
\caption[]{A possible identification of the X-ray sources
No. 5,7,8,9 from Hasinger et al.\, (1994). 
Also the positional deviations of the X-ray source No. 3 
from the galaxy are given. }
\begin{flushleft}
\begin{tabular}{lrrrr}
\hline\noalign{\smallskip}
X-ray  & Optical  &  $\Delta_{\alpha}$ & 
                     $\Delta_{\delta}$ & $\sigma_{\rm X-ray}$  \\ 
 source   & counterpart &    [\arcsec]   &   [\arcsec] & [\arcsec]   \\
\noalign{\smallskip}
\hline\noalign{\smallskip}
    5  & GYSB 32  & $-$4.6 & +0.4 &  \plm0.5    \\
    7  & AKO\,9    & $-$4.6 & +0.9 & \plm0.2    \\
    8  & GYSB 312 & $-$5.6 & $-$0.1 & \plm1.5   \\
    9  & DPF 2588 & $-$6.0 & +0.4 & \plm0.2    \\
\noalign{\smallskip}
\hline
\noalign{\smallskip}
Mean   &          &   $-$5.9 & +0.4 &        \\ 
$\sigma$ &        &   \plm0.7 & \plm0.4  &   \\
         &         &         &   &         \\
    3  & galaxy   &  $-$4.2 & +1.6 & \plm1.7  \\
\noalign{\smallskip}
\hline
\end{tabular}
\end{flushleft}
$\Delta_{\alpha}$ = Pos. difference (ROSAT minus ours) in 
      $\alpha$ \\  
$\Delta_{\delta}$ = Pos. difference (ROSAT minus ours) in
      $\delta$ \\
 $\sigma_{\rm X-ray}$ = Int. error of the X-ray position  \\
AKO = Number in Auri\`ere et al.\, (1989) \\
GYSB = Number in Guhathakurta et al.\, (1992) \\
DPF = Number in De~Marchi et al.\, (1993) \\
$\sigma$ = Standard deviation of the differences \\
\end{table}

\begin{table}
\caption[]{Alternative identification of the X-ray sources
No. 7,8,9,10 from Hasinger et al.\, (1994). }
\begin{flushleft}
\begin{tabular}{lrrrr}
\hline\noalign{\smallskip}
X-ray  & Optical  &  $\Delta_{\alpha}$ & 
                     $\Delta_{\delta}$ & $\sigma_{\rm X-ray}$  \\ 
 source   & counterpart &     [\arcsec]   &   [\arcsec] & [\arcsec]   \\
\noalign{\smallskip}
\hline\noalign{\smallskip}
    7  &  GYSB 172  & $-$0.6 & $-$3.3 &  \plm0.2 \\
    8  & GYSB 253 & $-$1.7 & $-$2.2 & \plm1.5  \\
    9  & AKO\,9 &   $-$1.6 & $-$4.4 & \plm0.2   \\
   10  & DPF 1286 &  $-$0.7 & $-3.8$  & \plm2.8    \\
\noalign{\smallskip}
\hline
\noalign{\smallskip}
 Mean  &          &   $-$1.2 & $-$3.4     &        \\
       &          &  \plm0.6 &    \plm0.9 &        \\
\noalign{\smallskip}
\hline
\end{tabular}
\end{flushleft}
$\Delta_{\alpha}$ = Pos. difference (ROSAT minus ours) in 
      $\alpha$ \\  
$\Delta_{\delta}$ = Pos. difference (ROSAT minus ours) in
      $\delta$ \\
 $\sigma_{\rm X-ray}$ = Int. error of the X-ray position \\ 
AKO = Number in Auri\`ere et al.\, (1989) \\
GYSB = Number in Guhathakurta et al.\, (1992) \\
DPF = Number in De~Marchi et al.\, (1993) \\
$\sigma$ = Standard deviation of the differences \\
\end{table}

\begin{table}
\caption[]{A third possible identification of the X-ray sources
No. 7,8,9,10 from Hasinger et al.\, (1994). }
\begin{flushleft}
\begin{tabular}{lrrrr}
\hline\noalign{\smallskip}
X-ray  & Optical  &  $\Delta_{\alpha}$ & 
                     $\Delta_{\delta}$ & $\sigma_{\rm X-ray}$  \\ 
 source   & counterpart &     [\arcsec]   &   [\arcsec] & [\arcsec]   \\
\noalign{\smallskip}
\hline\noalign{\smallskip}
    7  &  GYSB 241  & $-$6.2 & $-$3.1 &  \plm0.2 \\
    8  & E7 & $-$6.7 & $-$4.0 & \plm1.5  \\
    9  & V2 &   $-$6.6 & $-$3.7 & \plm0.2   \\
   10  & DPF 1206 &  $-$6.5 & $-6.0$  & \plm2.8    \\
\noalign{\smallskip}
\hline
\noalign{\smallskip}
 Mean  &          &   $-$6.5 & +4.2     &        \\
       &          &  \plm0.2 &    \plm1.3 &        \\
\noalign{\smallskip}
\hline
\end{tabular}
\end{flushleft}
$\Delta_{\alpha}$ = Pos. difference (ROSAT minus ours) in 
      $\alpha$ \\  
$\Delta_{\delta}$ = Pos. difference (ROSAT minus ours) in
      $\delta$ \\
 $\sigma_{\rm X-ray}$ = Int. error of the X-ray position \\ 
GYSB = Number in Guhathakurta et al.\, (1992) \\
E = Number in Edmonds et al.\, (1996) \\
DPF = Number in De~Marchi et al.\, (1993) \\
V2 = is the variable star from Paresce \& De Marchi (1994) \\
$\sigma$ = Standard deviation of the differences \\
\end{table}

One possibility for the determination of the constant
offset of the ROSAT positions would be the use of the older
Einstein X-ray position. As mentioned in Hasinger et 
al.\, (1994), the position of the X-ray source in 47\,Tuc 
from  the Einstein satellite (Grindlay et al.\, 1984) does 
not fit by its position to any of these sources detected by ROSAT.
However, we learned from earlier investigations that the X-ray 
positions from the Einstein satellite may have an error of about 
3$\arcsec$ (Geffert et al.\, 1989, Geffert et al.\, 1994).
The differences in position by the Einstein and ROSAT satellite 
may be also explained either by the constant positional shift
of the ROSAT positions or by a possible variability of the
X-ray source detected by the Einstein satellite. Taking
into account that the position of the Einstein satellite 
is a mean of five pointings, one may assume that this position
is the average of several sources, which were mixed and may
have varied during the observations. Therefore we have not
made use of the Einstein data in this paper.

\section {Identification of X-ray sources outside the central
          region of 47\,Tuc}

In the following we will discuss the identification of the
X-ray sources in the outer field of 47\,Tuc. An identified 
X-ray source in the outer region of 47\,Tuc would be
 important for the determination
of the constant positional shift of the ROSAT
positions.

\subsection {Identification of source No. 3}

All except one of the X-ray positions are located so close
to  the centre 
of 47\,Tuc, that an identification on the ESO Quick Blue 
Survey copy is not possible due to crowding problems. Only the X-ray source 3
lies sufficiently to the outside to allow an optical identification.
Given an error circle of about 10$\arcsec$ we found  
a faint galaxy on the ESO Survey  at the 
place of the X-ray  source No. 3. 
We have determined the optical position of the galaxy on two
ESO GPO plates with respect to 44 PPM stars. A position
of $\alpha_{2000}$ = 00$^h$23$^m$30$\fs$8 and $\delta_{2000}$ = 
$-$72$^{\gfull}$20$\arcmin$44$\gsecdot$0 was obtained, 
which shows an offset of $-$4$\gsecdot$2 
in $\alpha$ and +1$\gsecdot$6  in delta
(in the sense ROSAT position minus ours). Although the image
of the galaxy is quite faint on the GPO plates, the internal error
is with \plm 0$\gsecdot$4  much smaller than the internal error
of the ROSAT position, which amounts to \plm 1$\gsecdot$7.

\subsection {Identification of source No. 12}

The star HD 2072, which is according to Hasinger et al.\, (1994) 
the optical counterpart of source No. 12, is identical with
the star 366 905 of the PPM catalogue. Since the PPM position
is more precise than the position of the CDS used by
Hasinger et al.\, (1994), we have taken the PPM position
for the comparison with the ROSAT position.
The position difference between the X-ray source 
No. 12 and HD 2072 is +3$\gsecdot$2 in $\alpha$ 
and +1$\gsecdot$7 in $\delta$, 
which amounts to a total difference of 3$\gsecdot$3.

Because the positional differences between X-ray
and optical position of sources No. 3 and 12 are nearly in
opposite directions it is impossible that both 
identifications are true. It is
at this point impossible to judge which of the 
identifications is valid. Galaxies may be good X-ray
source candidates but also the explanation 
of Hasinger et al. (1994) that HD 2072 as a 
late type star is an X-ray emitter seems plausible.

\subsection {Identification of source No. 6}

Hasinger et al.\, (1994) have suggested that X-ray 
source No. 6 could be identified with CPD $-$72 35B.
However, the situation with CPD $-$72 35B is 
confusing. In the CPD star number $-$72 35 is designated as ``neb.``. 
The HD catalogue gives a cross identification of CPD $-$72 35
with HD 2051 but points out that star number HD 2051 is 47\,Tuc. On 
the other hand the Southern Double Star Catalogue of Innes (1927)
lists a double star with an angular separation of 6.21 arcsec 
and a position angle of 251$^{\circ}$, already found
by Herschel. This double star is also in the Hipparcos Input 
catalogue with the number HIC 1902 (CCDM 00241$-$7206).
Due to the uncertainty of the absolute position of
Herschel, we believe that the identification of
CPD $-$72 35 with Herschel's double star was a 
misidentification. Looking at our own CCD data we 
found the double star on our frames. The brighter 
component corresponds to the B8 III star already 
mentioned by Feast and Thackeray (1960) and which was 
found to be UV bright (de Boer \& van Albada 1976, de Boer 1985).
From our data we determined a position of
$\alpha_{2000}$ = 00$^h$23$^m$58$\fs$23 and $\delta_{2000}$ = 
$-$72$^{\gfull}$05$\arcmin$30$\gsecdot$1  
and a separation of 6.57 arcsec and a position angle 
of 247$\gfulldot$4 for this double star.  Since this position
is far away of any published X-ray position, we may rule 
out therefore an identification of the X-ray source 
No. 6 with any component of the double star.

\section {On the identification of the X-ray 
          sources in the core of 47\,Tuc}

In the following we look  for an identification of the 
X-ray sources under the assumption, that
most of the counterparts are objects from our Table 3.
However we cannot rule out that the majority of the optical 
counterparts of X-ray sources may be too faint for our observations or
look like normal stars.   Taking
into account a possible ROSAT shift of the X-ray position, we have
looked for all objects from our Table 3 within a radius 
of 10 arcsec of the X-ray source No. 7. The source No. 7 was 
chosen due to its small internal error and since the source
was seen in both observational runs of Hasinger et al.\, (1994).
9 candidates from Table 3 were found. For each candidate, the
difference between optical and X-ray position was used to 
shift the X-ray positions to the optical system. Then we looked 
in the error circle of each 
shifted X-ray position for objects from Table 3. Three possible
solutions were found, by which  4 X-ray sources could be identified 
with candidates from Table 3, two solutions containing the most
puzzling binary AKO\,9 as a candidate.

\subsection {Is AKO\,9 an optical counterpart of one of the central
             X-ray sources? }

AKO\,9 is historically  the candidate for the optical counterpart of the
X-ray position from EINSTEIN observations (Auri\`ere et al. 1989).
It was found as the hottest resolved object in the error box for
X0021.8-7221 during their 1986 July observations. Since then
it has been shown to have faded by about 2 mag in U (Bailyn 1990a).
AKO\,9  could be identified with DPF 2059 on the ESO 2.2m observations, 
but was not visible on 1990 NTT observations, which confirms a 2 mag fading 
(Auri\`ere et al 1994). 
Now, Meylan et al (1996), using the HST, observed it flaring, its 
UV brightness increasing by about 3 magnitudes in 2 hours. Edmonds 
et al.\, (1996) found AKO\,9 (their variable 11) as an eclipsing binary. 
AKO\,9 is thus a peculiar binary which could be an X-ray source.
The position of AKO\,9 is within 5 $\arcsec$ of the two ROSAT sources
No. 7 amd No. 9 from Hasinger et al.\ (1994).
The known properties of AKO\,9 have been analysed by Minniti et al.\, (1996)
who, however, could not derive its true nature since the observed flare in the
UV has an unusual shape. The possibilities explored so far to explain the
flare involve four types of objects (RS CVn, Cataclysmic Variable, soft-X-ray 
transient, nova) which are classically associated to X-ray sources which 
could be luminous enough to be detected with ROSAT at the distance of 47\,Tuc.  

Our first solution is presented in Table 4.
This solution is in good agreement with the identification
of the X-ray source No. 3 with the galaxy, as mentioned in chapter 4.
Table 4 lists the identification and their $O-C$ data of 
this solution. The positional error of the ROSAT position of 
source 3 of 1$\gsecdot$7  may explain the differences in column 3 and 4  
between source 3 and the other sources.

On the other hand we arrive at an  alternative solution taking
into account the identification of AKO\,9 with X-ray source No. 9. 
This would lead also to three other identifications  
of X-ray sources with blue stragglers. The same possibility was also 
taken into account by Meylan et al.\, (1996) and Minniti et al.\,
(1996). 
The identifications with their differences between 
the X-ray and optical positions are given in Table 5.
We wish to point out that both solutions are of identical 
value from astrometric point of view. Only the probable
identifications of the X-ray source 3 would lead us
to a preferred choice of the solution in Table 4.
As to the solution in Table 5, the fact that X9 and
AKO\,9 are both transient objects with compatible positions 
is a favourable argument. It has to be noticed that  
AKO\,9 could be associated with a ROSAT X-ray source 
even if it was not the case for any of the blue stragglers.

\subsection {Variable stars as counterparts of the X-ray sources}

Paresce et al.\, (1992) have proposed V1, a cataclysmic variable 
in the core of 47\,Tuc, as the counterpart of  one of the X-ray sources. 
V2, a dwarf nova, was discovered later (Paresce \& De Marchi, 1994). 
V1 corresponds to DPF 2213 (Paresce et al.\, 1992) and V2 is located 
near to DPF 2184 (this follows from comparison of the data and charts
of De~Marchi et al. (1993) with the corresponding chart of
Paresce \& De Marchi, (1994)). Even when taking a possible shift of 
the ROSAT positions into account, it is not possible
with the given relative positions of the identified objects
to identify both variable stars together with any two 
ROSAT positions. 

Recently Gilliland et al.\, (1995) and 
Edmonds et al.\, (1996) using HST observations 
have detected 13 variable stars in the central
66$\arcsec$ $\times$ 66$\arcsec$ region. Since their  positions were given
in the same system as those from Guhathakurta et al. (1992),
we were able to determine the positions of the 
variable stars. Assuming a constant shift of the
X-ray positions of Hasinger et al.\, (1994) can provide
coincidences of optical and X-ray positions for 
only two objects, and then not unique ones but for several pairs
of stars. Edmonds
et al. (1996) are in favour of a solution  which is in
agreement with the identification of HD 2072 with the
Source X12 from Hasinger et al.\, (1994). The astronomical
data of their candidates give additional arguments for  this solution.
However, their identification of the variable 7 
with the source X10 is astrometrically weak due to an error circle 
of 2$\gsecdot$8 of the X-ray position. In addition,
three of the central X-ray sources would then remain as
not identified. 

Taking blue stragglers and variable stars into
account, we found an additional solution for the identification
of four central X-ray sources with objects from Table 3.
These identifications and their astrometric
differences are presented in Table 6. 

\section{Discussion of the different solutions}

The three solutions presented above could give for most of
the X-ray sources in the core of 47\,Tuc an optical counterpart. 
Moreover AKO\,9, which is, due to its blue colour and variability, 
a very probable candidate of an X-ray source, is included
in the identification. One solution is in agreement
with the identification of one of the outer X-ray sources with a galaxy. 
This solution from Table 4
is our preferred solution. On the other hand, the fact that we
have three nearly identical identifications, from astrometric point 
of view, for the central sources points to the fact that 
we have also superpositions by chance. 
   
The solution of Edmonds et al.\, (1996) is compatible
with transients in quiescence or magnetic cataclysmic variables. However, it
concerns only two faint variable stars for which even the colour could not
be measured. This also would imply that the non identified objects
are very faint. Also, in this case,  a larger number of possible 
identifications of one or two central X-ray sources exist. Nevertheless, 
one of the main interests of the solution of Edmonds et al.\, (1996)
is to be compatible with HD 2072 being associated with X12. 

ROSAT observed two blue stragglers in open clusters associated with X-ray sources: one
in M67 and one in  NGC 752 (Belloni \& Verbunt, 1996). However, these are
faint sources (around 10$^{30}$ erg s$^{-1}$) and generally in 47\,Tuc 
and other globular (and open) clusters we see many blue 
stragglers, which do
not show X-ray emmission. Our question must therefore be,
whether the optical counterparts of the X-ray sources in 
47\,Tuc may have the photometric characteristics of
blue stragglers. This would imply that we are dealing with objects
which by chance have similar photometric properties as
classical blue stragglers.
As presented in our introduction, the X-ray sources in the
central field of 47\,Tuc are puzzling since their X-ray luminosities are of
an order of 1 magnitude higher
than those of cataclysmic variables but much fainter than those of normal
X-ray binaries. Hasinger et al.\, (1994) propose that these objects
belong to the class of transient LMXBs in quiescence,
similar to those objects in the galactic disk, which were 
found before by Verbunt et al.\, (1994). These objects are expected to
appear very faint in the visible at the distance of 47\,Tuc 
(Auri\`ere et al 1989).
 One possibility in the present context 
could be that the 47\,Tuc X-ray sources are active binaries 
as those discussed in Bailyn (1990b) and as could be AKO\,9,
which could mimic photometric properties of BSs.
A safe identification will only be possible
either by improvement of the absolute optical and X-ray
positions or by coincident brightness variations in the
optical and in the X-ray spectral range. 
In addition, it would be good to obtain spectra of the 
galaxy and of HD2072 in order to determine their nature and 
distance to test the hypothesis that one of
them could be an X-ray source.

\section{Search for optical counterparts of the millisecond
         pulsars in 47\,Tuc}

11 millisecond  pulsars have been found in 47\,Tuc but for only
two, 47\,Tuc C and 47\,Tuc D, accurate radio positions (better 
than 100 mas) have been determined (Robinson et al.\, 1995).
Unfortunately 47\,Tuc C is located outside 
all of our CCD or HST fields. The position of pulsar 47\,Tuc D 
is only within the field of our NTT observations. At the position of 
the pulsar we found a star with $B$=14.91 and $U-B$=0.32 at an offset
of 1$\gsecdot$3 and another star with $B$=18.08 and $U-B$=0.20 with an
offset of 1$\gsecdot$4  relative to the pulsar position. Since our positions
are based on the PPM system, the positional differences of both stars 
from the radio position of the pulsar seem to be too large to
propose an identification of one of these stars with the
pulsar. However, a final decision will only  be possible
with the results of the Hipparcos mission.

\section{Summary}

We have determined optical positions of blue and variable stars 
in the centre of 47\,Tuc with an internal accuracy of better than
100 mas. These data were used for a reinvestigation 
of the identification of the X-ray sources in 47\,Tuc 
found by Hasinger et al.\, (1994).
Under the assumption that the optical counterparts of the X-ray 
sources in 47\,Tuc are stars from our Table 3, we present three
solutions, which give optical counterparts of four central
X-ray sources; one solution is in agreement with our
proposition of identification of X-ray source No. 3, in the
outer part of the cluster, with a galaxy.

Nevertheless, the solutions with blue stragglers as
optical counterparts contain two problems.
In the optical most of the candidates cannot be distinguished
from classical blue stragglers, which are not generelly
bright X-ray sources.
In addition we 
cannot rule out that these identifications are 
only by chance. Advantages for our solutions are
to give counterparts for four central sources
and to include in two cases the most puzzling
binary of the 47\,Tuc core, AKO\,9.
If all of our solutions are wrong, then we may conclude
that the majority of the true sources were too faint to be 
identified as blue or variable objects by the first HST
observations.

\acknowledgements 
We thank G. De~Marchi and C. Lauzeral for 
providing us with their data in electronic
readible form. We are also indebted to H.-J. Tucholke
(Bonn) for lending us two deeper plates of 47\,Tuc for measuring 
the position of the galaxy in chapter 5. We are grateful to
H.M. Johnston, G. Meylan, D. Minniti, F. Verbunt, 
G. Hasinger and P. Edmonds
for useful discussions and preprints.

\end{document}